\documentclass{jltp}
\usepackage{amssymb}

\runninghead{W. J. Mullin, M. Holzmann, and F. 
Lalo\"{e}}{Hohenberg Theorem for a Generalized Bose-Einstein 
Condensation}

\title{Validity of the Hohenberg Theorem for a Generalized Bose-Einstein
Condensation in Two Dimensions}
\author{W. J. Mullin$^{a}$, M. Holzmann$^{b}$, and F. Lalo\"{e}$^{b}$}
\address{$^{a}$Physics Department, Hasbrouck Laboratory\\
University of Massachusetts Amherst, MA 01003 USA\\
$^{b}$LKB, D\'{e}partement de Physique de l'ENS\\
24 rue Lhomond  75005 Paris, France\\
}

\begin{document}
\maketitle

\begin{abstract}
\noindent Several authors have considered the possibility of a
generalized Bose-Einstein condensation (BEC) in which a band of low
states is occupied so that the total occupation number is macroscopic,
even if the occupation number of each state is not extensive.  The
Hohenberg theorem (HT) states that there is no BEC into a single state
in 2D; we consider its validity for the case of a generalized
condensation and find that, under certain conditions, the HT does 
{\em not}
forbid a BEC in 2D. We discuss whether this situation actually occurs
in any theoretical model system.

\end{abstract}


\section{INTRODUCTION}

The Hohenberg theorem\cite{Hohen} (HT) provides the general statement
that Bose-Einstein condensation (BEC) cannot occur in a
two-dimensional system.  In this analysis a condensation implies
extensive occupation of a single state of the system, that is, a
density of particles of order $N/V,$ (where $N$ is the number of
particles in the system and $V$ the volume of the system) in the
thermodynamic limit.  Recently however there has been renewed interest
in the possibility of a ``smeared,'' ``fragmented,'' or
``generalized'' BEC, in which some finite band of states, rather than
a single state, is occupied.  Nozi\`{e}res and Saint
James\cite{StJames} and more recently Nozi\`{e}res\cite{Noz}, using a
Hartree-Fock approximation at zero temperature, showed that repulsive
interactions favor single-state occupancy.  There have been earlier
discussions of generalized BEC in the literature, for example, by
Girardeau%
\cite{Gir}, Luban\cite{Lub} and Van den Berg \emph{et al}\cite{VdBerg}. Van
den Berg and Lewis\cite{VdBLew} have even presented a non-interacting model
in which an extensive BEC occurs in each state of a band of momentum states.
Ho and Yip\cite{HoYip} discuss fragmentation in a recent paper and claim
that the spin-1 Bose gas\cite{Bigelow} is an example of the occurrence of
this phenomenon.

In two dimensions (2D), where the HT applies, there is no single-state
BEC, but there is a Kosterlitz-Thouless\cite{KT} transition to a superfluid
state. However, one can ask whether there could be a generalized transition,
and, if so, whether it is related to the KT transition, or whether the
generalized transition is also forbidden by the HT.

If the generalized BEC is to be possible, we might envision different 
forms of it.  One possibility (which we call ``fragmentation'') is 
that each of a finite number of states is extensively occupied  (occupation 
proportional to $N$) and that the sum of all the particles in such a 
condensate is still of order $N$.  Another interesting possibility 
(here termed ``smearing'') is that no single state is extensively occupied, 
but that the sum of the occupation numbers of all the states in a band 
is extensive.  For example, one might have $O(\sqrt{N})$ states each 
occupied by $O(\sqrt{N})$ particles so that the total number of 
particles is of order $N.$ The question is whether the HT forbids either 
of these kinds of generalized condensation in 2D.

The generalized condensation studied in Ref.\ 
\cite{VdBLew} 
occurs in 2D, which seems to violate the HT, but there are some subtle 
aspects that need to be considered to see how this case fits into the 
general picture.  The example given there consists of particles moving 
freely in one dimension and bound harmonically in the other.  The 
condensation occurs in the lowest harmonic band of states.  In a 
similar problem, free particles trapped in a 2D harmonic potential are 
known to have a BEC, but this arises because there is a ``loophole'' 
in the HT, due to of the inhomogeneous potential.  One can 
show\cite{Mul1} that the HT does indeed apply to the case of trapped 
inhomogenous fluids, but that its application requires that the 
particle density be everywhere bounded.  When Bose condensed, the 
ideal gas in a trap has an integrable singularity in its density in 
the thermodynamic limit, and thus does not fall under the conditions 
covered by the HT. However, as soon as repulsive hard-core 
interactions are turned on, this singularity disappears and the HT\ 
applies.\cite{Mul1} Similarly, the example of Ref.\ 
\cite{VdBLew} is not a case of a violation of the HT\ because of 
fragmenting of the condensate, but rather a case where the HT does 
not apply because one has an ideal gas in a trap.

The situation we will consider here is a homogeneous system (particles in a
box) in which we suppose that a generalized BEC occurs as mentioned above,
namely, a narrow band of momentum states occupied either extensively or
non-extensively, but with the sum of their occupation numbers extensive.

\section{DERIVATION}

We consider a homogenous system of $N$ particles. We follow the derivation
of the Hohenberg theorem due to Chester.\cite{Chester} Start with the
Bogoliubov inequality given by 
\begin{equation}
\left\langle \frac{1}{2}\left\{ A,A^{\dagger }\right\} \right\rangle
\geqslant \frac{kT\left| \left\langle \left[ C,A\right] \right\rangle
\right| ^{2}}{\left\langle \left[ \left[ C,H\right] ,C^{\dagger }\right]
\right\rangle },
\end{equation}
where $\left\langle ...\right\rangle $ means thermal average, $A$ and $C$
are operators, and $H$ is the system Hamiltonian. Choose $A=a_{\mathbf{p}}a_{%
\mathbf{p}+\mathbf{k}}^{\dagger },$ and $C=\sum_{\mathbf{q}}a_{\mathbf{q}%
}^{\dagger }a_{\mathbf{q}+\mathbf{p}},$ where $a_{\mathbf{p}}^{\dagger }$
creates a particle with momentum $\mathbf{p}.$ $C,$ being the Fourier
transform of the density operator, commutes with the interaction potential
energy in $H$, so we need only to consider commutators with the kinetic
energy. Upon carrying out these commutators we find 
\begin{eqnarray}
\left\langle \frac{1}{2}\left\{ A,A^{\dagger }\right\} \right\rangle  &=&n_{%
\mathbf{p}}n_{\mathbf{p}+\mathbf{k}}+\frac{1}{2}(n_{\mathbf{p}}+n_{\mathbf{k}%
})+\delta _{\mathbf{k},0}(n_{\mathbf{p}}+1), \\
\left\langle \left[ C,A\right] \right\rangle  &=&n_{\mathbf{p}}-n_{\mathbf{p}%
+\mathbf{k}}, \\
\left\langle \left[ \left[ C,H\right] ,C^{\dagger }\right] \right\rangle  &=&%
\frac{\hbar ^{2}}{2m}\sum_{\mathbf{q}}\left[ (\mathbf{k}+\mathbf{q})^{2}+(%
\mathbf{k}-\mathbf{q})^{2}-2\mathbf{q}^{2}\right] n_{\mathbf{q}}  \nonumber
\\
&=&\frac{\hbar ^{2}k^{2}}{m}\sum_{\mathbf{q}}n_{\mathbf{q}}=N\frac{\hbar
^{2}k^{2}}{m},
\end{eqnarray}
where $n_{\mathbf{p}}$ is the thermal average number of particles in state $%
\mathbf{p}$. The inequality becomes 
\begin{equation}
n_{\mathbf{p}}n_{\mathbf{p}+\mathbf{k}}+\frac{1}{2}(n_{\mathbf{p}}+n_{%
\mathbf{k}})+\delta _{\mathbf{k},0}(n_{\mathbf{p}}+1)\geqslant \frac{kTm}{%
N\hbar ^{2}k^{2}}(n_{\mathbf{p}}-n_{\mathbf{p}+\mathbf{k}})^{2}.
\label{basic}
\end{equation}

We assume that there are $M_{c}$ condensed states (our generalized 
condensate band) containing $ N_{0}$ particles with $N_{0}=O(N)$, and 
that these states are clustered in momentum space around 
$\mathbf{k}=0$ in a circle out to radius $p_{c}$.  The non-condensed 
states are in the region beyond $p_{c}.$ We assume that the condensed 
states each have occupation that, while not necessarily extensive, still 
greatly exceeds that of any non-condensed state.  We have then
$
\sum_{p<p_{c}}1=M_{c},\hspace{0.2in}\sum_{p<p_{c}}n_{\mathbf{p}}=N_{0}.$
If we change the sum to an integral, the first of these equations becomes 
\begin{equation}
M_{c}=\sum_{\mathbf{p}=0}^{p_{c}}1=\frac{L^{2}}{2\pi }\frac{p_{c}^{2}}{2},
\end{equation}
where $L$ is the dimension of the box and $n=N/L^{2}$ is the density. Thus 
\begin{equation}
p_{c}=c_{1}\left( \frac{M_{c}}{N}\right) ^{1/2}n^{1/2},
\label{pc}
\end{equation}
where $c_{1}$ is a constant of order unity.

Sum both sides of Eq.(\ref{basic}) on $\mathbf{k}$ over the range $
2p_{c}<k<p_{m}$, where the upper limit $k_{m}$ satisfies $k_{m}=c_{2}n^{1/2}$
with $c_{2}$ a constant. Below we will see why we take the minimum $k$ value
as twice $p_{c}$ rather than just $p_{c}.$ We have 
\begin{equation}
n_{\mathbf{p}}\sum_{\mathbf{k}}n_{\mathbf{p}+\mathbf{k}}+\frac{1}{2}\sum_{%
\mathbf{k}}(n_{\mathbf{p}}+n_{\mathbf{k}})\geqslant \frac{kTm}{N\hbar ^{2}}
\sum_{\mathbf{k}}\frac{(n_{\mathbf{p}}-n_{\mathbf{p}+\mathbf{k}})^{2}}{k^{2}}.
\label{basic2}
\end{equation}
We assume that $\mathbf{p}$ is in the range of condensed states so that the
second term in the second sum on the left is much smaller than the first and
can be neglected. The remaining term in that sum is 
\begin{equation}
n_{\mathbf{p}}\sum_{\mathbf{k}}1=\frac{L^{2}}{2\pi }%
\int_{2p_{c}}^{p_{m}}dkk=n_{\mathbf{p}}\frac{N}{n(4\pi )}%
(p_{m}^{2}-4p_{c}^{2})=Nn_{\mathbf{p}}\gamma,
\end{equation}
where $\gamma =(c_{2}-4c_{1}\frac{M_{c}}{N})/(4\pi )$ is a constant of order
unity. The first sum on the left of Eq.(\ref{basic2}) is over only a portion
of momentum space and results in a fraction of the total particle number.
Increasing the sum to the full number of particles just amplifies the
inequality. We then can write $n_{\mathbf{p}}N(1+\gamma )$ for the left side.

The sum on the right side of Eq.(\ref{basic2}) has the vector
$\mathbf{k}+\mathbf{p}$ extending outside the condensate range because
of the restriction that $k>2p_{c}.$ If
we had defined the minimum value of $k$ as just $p_{c}$ then, by putting $\mathbf{k%
}$ and $\mathbf{p}$ in opposite directions, the sum could extend back
into the condensate circle.  As it is we can neglect
$n_{\mathbf{p}+\mathbf{k}}$ relative to $n_{\mathbf{p}}$ on the right
side of the equation to give simply
$n_{\mathbf{p}}^{2}\sum_{\mathbf{k}}1/k^{2}.$ The sum is evaluated by
doing the appropriate integral: 
\begin{equation}
\sum_{\mathbf{k}}\frac{1}{k^{2}}=\frac{N}{n(2\pi )}\int_{2p_{c}}^{p_{m}}dk%
\frac{1}{k}=c_{3}\frac{N}{n}\ln \left( \frac{p_{m}}{2p_{c}}\right).
\end{equation}
The result is 
\begin{equation}
n_{\mathbf{p}}N(1+\gamma )\geqslant n_{\mathbf{p}}^{2}\frac{kTm}{N\hbar ^{2}}
c_{3}\frac{N}{n}\ln \left( \frac{p_{m}}{2p_{c}}\right) 
\end{equation}
or 
\begin{equation}
\frac{n_{\mathbf{p}}}{N}\leqslant \frac{\hbar ^{2}n}{mkTc_{3}}\frac{1+\gamma 
}{\ln \left( \frac{p_{m}}{2p_{c}}\right) }.
\end{equation}

Now sum this over all states in the condensate circle to give 
\begin{equation}
\frac{N_{0}}{N}\leqslant \frac{\hbar ^{2}n}{mkTc_{3}}\frac{\left( 1+\gamma
\right) }{\ln \left( c_{4}\sqrt{\frac{N}{M_{c}}}\right) }M_{c}.
\end{equation}
where we have used the fact that
\begin{equation}
\frac{p_{m}}{2p_{c}}=\frac{c_{2}\sqrt{n}}{2c_{1}\sqrt{n\frac{M_{c}}{N}}}
=c_{4}\sqrt{\frac{N}{M_{c}}}. 
\end{equation}

If $M_{c}=1$, as in the usual case, then, in the thermodynamic limit,
the right side goes to zero as $1/\ln (N)$ and there is no BEC. However, if $
M_{c}$ $=O(N^{\nu })$, where $\nu $ is any power not equal to zero, then the
inequality becomes 
\begin{equation}
\frac{N_{0}}{N}\leqslant \frac{\hbar ^{2}n}{mkT}\frac{N^{\nu }}{\ln \left(
N^{(1-\nu )/2}\right) }.
\end{equation}
Now the right side \emph{diverges} as $N\rightarrow \infty.$ For example,
if $\nu =\frac{1}{2}$ then the right side is of order $\sqrt{N}/\ln (N),$
which clearly diverges. In this case we can no longer rule out the
possibility of BEC by this argument. 

\section{DISCUSSION}

Our derivation does not \emph{prove} that BEC actually happens in a
generalized mode. Moreover, it also does not insure that there might
not be \emph{another} more powerful derivation of the HT\ that outlaws a
generalized condensation in 2D.  In the case of a fragmented
condensate (as in Ref.\ 
\cite{VdBLew}), one can have only a
finite number, $M_{c}$, of states each containing $O(N)$ particles so
that $M_{c}$ is $O(1)$ and \emph{there can be no fragmented BEC in a
2D homogeneous system}.  The fragmented case of Ref.\
\cite{VdBLew} is ``tainted'' by the fact that the HT conditions
for the inhomgeneous case are violated by a divergent density of
particles so that it is not a counter-example to the HT case that we
discuss in this paper.  A 2D version of the spin-1 gas, claimed in
Ref.\ 
\cite{HoYip} to be an example of fragmentation in 3D,
would not fall directly under our derivation because it is a
condensation in \emph{spin} space, not in the momentum space of
our derivation.  (The spin-1 Bose system appears to be a case
that violates the claim of Refs.\ 
\cite{StJames} and
\cite{Noz} that a single-state BEC was favored over generalized
BEC. However, that discussion was based on use of mean-field theory
and the spin-1 system analyses go beyond mean-field approximations.)

On the other hand, the analysis above leaves open the possibility of a 
smeared BEC in 2D. In this case, $N^{\nu} (\nu<1)$ states each 
containing only a non-extensive number of particles constitute a 
condensate as a unit.  Note that the width of this band of states, as 
given in Eq.(\ref{pc}), tends to zero as $N$ increases, so 
one still gets a $\delta$-function occupation in the thermodynamic 
limit.

A question yet to be answered is whether the KT transition is related in
some way to a generalized BEC. The analysis of Popov\cite{Popov} of the KT state
at low temperature involves what he calls a ``bare'' condensate spread over
a set of states up to some cut-off $k_{0}.$ He says ``the particles with
momenta small compared with the faster particles behave like a condensate.''
However, it is not clear to us at this time whether this situation would
qualify as a generalized condensate in the sense of our theorem above.


\section{ACKNOWLEDGMENTS}

The Laboratoire Kastler-Brossel is Unit\'{e} Associ\'{e}e au CNRS (UMR8552) et
\`{a} l'Universit\'{e} Pierre et Marie Curie. WJM would like to thank Ecole
Normale Sup\'{e}rieure, where some of this research was carried out, for a
travel grant and for its excellent hospitality.

\end{document}